\documentclass{article}
\usepackage{amssymb,epsfig,graphicx}
\usepackage{amsmath, graphics}
\newcommand{\be}{\begin{equation}}
\newcommand{\ee}{\end{equation}}
\newcommand{\bea}{\begin{eqnarray}}
\newcommand{\eea}{\end{eqnarray}}
\newcommand{\barr}{\begin{array}}
\newcommand{\earr}{\end{array}}

\newtheorem{prop}{Proposition}[section]
\newtheorem{rem}{Remark}[section]

\newcommand{\bpar}{\be \left\{ \begin{array}{lll}}
\newcommand{\epar}{ \end{array}\right. \ee }
\newcommand{\eparn}{ \end{array} \right.}

\newcommand\leftmat{\left(\begin{array}{cc}}
\newcommand\rightmat{\end{array}\right)}
\newcommand\leftvec{\left(\begin{array}{c}}
\newcommand\rightvec{\end{array}\right)}
\newcommand\re{{\rm e}}
\begin{document}
\bibliographystyle{plain}

\vspace{3mm}
\begin{center}
\Large
{\bf The generalised Dirichlet to Neumann map for moving initial-boundary value problems}

\vspace{3mm}

\large
A.S. Fokas$^{*}$ and B. Pelloni$^{**}$

\vspace{5mm}

*Department of Applied Mathematics and Theoretical Physics

Cambridge University

Cambridge CB3 0WA, UK.

{\em t.fokas@damtp.cam.ac.uk}

\vspace{2mm}

**Department of Mathematics

University of Reading

Reading RG6 6AX, UK

{\em b.pelloni@rdg.ac.uk}

\vspace{3mm}
\today
\end{center}
\normalsize

\begin{abstract}
    We present an algorithm for characterising the generalised Dirichlet 
    to Neumann map for moving initial-boundary value problems.
    This algorithm  is derived by combining the so-called global 
    relation, which couples the  initial and boundary values 
    of the problem, with a new method for inverting certain one-dimensional integrals. 
    This new method is based on the spectral analysis of an 
    associated ODE and on the use of the d-bar formalism. 
    As an illustration, the Neumann boundary value for the linearised 
    Schr\"odinger equation is determined in terms of the Dirichlet boundary 
    value and  of the  initial condition. 
\end{abstract}

\section{Introduction}
\begin{figure}
    \begin{center}
	    \includegraphics[width=8cm]{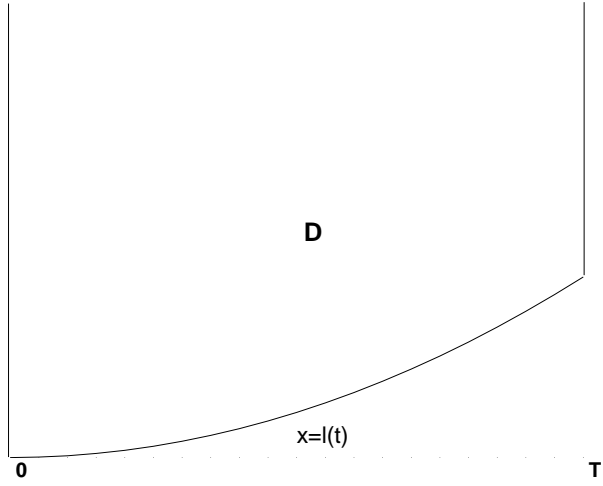}
    \caption{The domain ${\bf D}$ in the $(x,t)$ plane}
    \label{fig1}
    \end{center}
    \end{figure}

We present a methodology for characterising the generalised Dirichlet 
    to Neumann map for linear 
evolution PDEs posed on domains whose boundary varies with time. 
Consider, as an example, the following domain in the $(x,t)$ 
plane, see figure \ref{fig1}:
\be
{\bf D}:\quad 0<t<T,\quad  l(t)<x<\infty,
\label{Ddef}
\ee
where $T$ is a positive constant and  $l(t)$ is a given monotonic, 
twice differentiable function. For economy of presentation we assume 
that $l(t)$ satifies the following:
$$
l(t)\in{\bf C}^{2}[0,T], \quad l''(t)>0,\quad l(0)=0.
$$

Let the scalar complex-valued function $q(x,t)$ satisfy the boundary 
value problem
\bea
&& iq_{t}+q_{xx}=0,\quad (x,t)\in{\bf D},
 \nonumber \\
&&q(x,0)=q_{0}(x),\;\; 0<x<\infty; \quad q(l(t),t)=f_{0}(t),\;\;0<t<T.
\label{lsc}
\eea
We assume that the initial  condition $q_{0}(x)$ is a sufficiently smooth 
function, decaying as $x\to~\infty$, that the boundary condition 
$f_{0}(t)$ is sufficiently smooth, and that these two functions are 
compatible at the origin, $q_{0}(0)=f_{0}(0)$.

There exist several different integral 
representations for the solution of initial-boundary value 
problems such as those defined by 
(\ref{lsc}). 
These include the representation constructed by using the classical 
Fourier transform in the $x$ variable, as well as the novel 
representations presented in \cite{prl, fpel}. However, these 
representations are {\em not} effective, because they involve 
{\em unknown} boundary values. For example, for the initial-boundary value 
problem (\ref{lsc}) these representations involve the unknown function $q_{x}(l(t),t)$.

Thus, the key issue is the characterisation of the {\em generalised 
Dirichlet to Neumann map}, namely the characterisation of the unknown 
boundary values in terms of the given initial and boundary conditions.
For the initial-boundary value problem 
(\ref{lsc})  this means computing  $q_{x}(l(t),t)$ in terms of 
$\{q_{0}(x), f_{0}(t)\}$.

Here we present an algorithm for constructing the generalised 
Dirichlet to Neumann map. For the initial boundary value problem 
(\ref{lsc})  this algorithm yields the unknown function 
$q_{x}(l(t),t)=f_{1}(t)$ through the solution of the 
Volterra integral equation
\be
f_{1}(t)=\frac{2}{3\pi}
\int_{0}^{t}J(s,t)f_{1}(s)ds+
\label{volterra1} \ee
$$+\frac{2}{3\sqrt{\pi}}\re^{-i\pi/4}
% \Gamma\left(\frac{1}{2}\right)
\left[
\int_{0}^{t}\frac{\re^{i\frac{(l(t)-l(s))^{2}}{4(t-s)}}}{\sqrt{t-s}}f_{0}'(s)ds
+\frac{1}{\sqrt{t}}
\int_{0}^{\infty}\re^{\frac{i(l(t)-x)^{2}}{4t}}q_{0}'(x)dx\right],
$$
for $0<t<T$, 
% and the constant $A$ given by
% \be
% A=\frac{4}{3\pi}\re^{-i\pi/4}\Gamma\left(\frac{1}{2}\right)
% \label{gdef}
% \ee
% ($\Gamma$ is the usual Gamma function), 
where the kernel 
$J(s,t)$ is defined by
\be
J(s,t)=\frac{l(s)-l(t)}{2(s-t)}\left\{\int_{l'(s)/2}^{\infty}\re^{ik^{2}(s-t)-ik(l(s)-l(t))}dk-
i\,G(s,t)\int_{0}^{\infty}\re^{-ik^{2}(s-t)-k(s-t)\left[l'(s)-\frac{l(s)-l(t)}{s-t}\right]}dk\right\},
\label{Jdef1}\ee
and 
\be
G(s,t)=\re^{i\frac{l'(s)^{2}}{4}(s-t)-i\frac{l'(s)}{2}(l(s)-l(t))}.
\label{Gst}
\ee

We note the both integrals on the right hand side of equation 
(\ref{Jdef1}) are well defined. In particular, the second integral 
involves an exponential which {\em decays} as $k\to\infty$. Indeed, 
the real part of the exponent is 
$$
-k(s-t)\left(l'(s)-\frac{l(t)-l(s)}{t-s}\right)=-k(s-t)(l'(s)-l'(\sigma)),\quad 
s\leq \sigma\leq t.
$$
Recalling that $l''(t)>0$, it follows that $l'(s)-l'(\sigma)\leq 0$, 
and since 
$k\geq 0$ and $s-t\leq 0$, the real part of the exponent is negative.

% Both these integrals can be expressed in terms of the error function 
% $Erf$. 

\smallskip
The algorithm involves three steps:

{\bf 1.} {\em Assuming that the solution exists, derive the so-called global 
relation,  namely the relation that couples the initial condition 
with all boundary values}. This step is elementary and it involves 
only writing the given PDE in a proper divergence form, and applying  
Green's theorem. For example, equation (\ref{lsc}) can be rewritten in 
the form
\be
\left[\re^{-ikx+ik^{2}t}q(x,t)\right]_{t}-\left[\re^{-ikx+ik^{2}t}
(iq_{x}(x,t)-kq(x,t)\right]_{x}=0,
\quad k\in\mathbb C.
\label{lscdiv}
\ee
Then an application of Green's theorem in the domain ${\bf D}$ yields
$$
i\int_{0}^{T}\re^{ik^{2}s-ikl(s)}q_{x}(l(s),s)ds=\int_{0}^{T}\re^{ik^{2}s-ikl(s)}(k-l'(s))f_{0}(s)ds
$$
\be+
\hat{q}_{0}(k)
-\re^{ik^{2}T}\int_{l(T)}^{\infty}\re^{-ikx}q(x,T)dx,\quad {\rm 
Im}(k)\leq 0.
\label{ex1gr}
\ee

\begin{figure}[h]
    \begin{center}
 	\setlength{\unitlength}{0.04in}
	\begin{picture}(90,40)(-50,6)
      \thinlines
   \put(-10,35){\vector(-1,0){25}} 
   \put(-8,35){\line(1,0){1}}
   \put(-4,35){\line(1,0){1}}
   \put(0,35){\line(1,0){1}}
   \put(4,35){\line(1,0){1}}
   \put(8,35){\line(1,0){1}}
   \put(12,35){\line(1,0){1}}
   \put(16,35){\line(1,0){1}}
   \put(20,35){\line(1,0){1}}
   \put(23,34){$k_{I}=0$}
    \put(-35,35){\line(-1,0){25}}
    \put(-10,35){\line(0,-1){30}}
    \put(-10,15){\vector(0,1){5}}
    \put(-15,37){$k_{R}=\frac{l'(t)}{2}$}
%    %%
%    \put(10,35){\line(1,0){2}}
%       \put(14,35){\line(1,0){2}}
%       \put(18,35){\line(1,0){2}}
%       \put(22,35){\line(1,0){2}}
%       \put(26,35){\line(1,0){2}}
%       \put(30,35){\line(1,0){2}}
%       \put(34,35){\line(1,0){2}}
%       \put(38,35){\line(1,0){2}}
%       \put(42,35){\line(1,0){2}}
%       \put(46,35){\line(1,0){2}}
%       \put(50,35){\line(1,0){2}}
%       \put(54,35){\line(1,0){2}}
%      \put(10,35){\line(5,-1){45}}
%      \put(33,30.8){\vector(1,-1){1}}
%      \put(10,35){\line(0,-1){1}}
%      \put(10,32){\line(0,-1){1}}
%      \put(10,29){\line(0,-1){1}}
%      \put(10,26){\line(0,-1){1}}
%      \put(10,23){\line(0,-1){1}}
%      \put(10,20){\line(0,-1){1}}
%      \put(10,17){\line(0,-1){1}}
%      \put(10,14){\line(0,-1){1}}
%      \put(10,11){\line(0,-1){1}}
%      \put(10,8){\line(0,-1){1}}
%      \put(7,37){$l'(t)/2$}
%      \put(45,23){$\Gamma(t)$}
     \end{picture}
     \caption{The curve $\partial\Omega_{2}^{-}(t)$ 
     in the $k$ plane }
% \hspace{20pt}(2b) A 
%     curve $\Gamma(t)$}
 \label{fig2a} 
 \end{center}
     \end{figure}
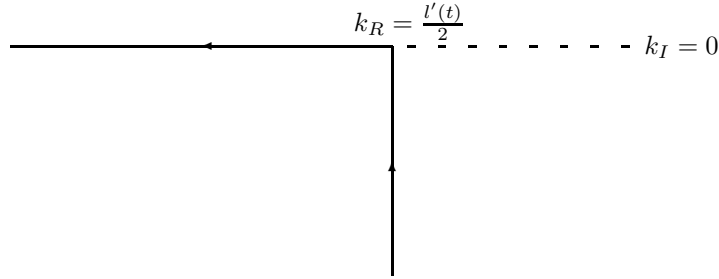

\smallskip
{\bf 2.} {\em Consider the integral involving the unknown boundary 
values and derive  a general formula for the inversion of this type of 
integrals.} 
This step involves the {\em spectral analysis} of an 
appropriate ODE. 
For example, for the initial boundary value problem (\ref{lsc}), this 
 ODE is
\be
\mu_{t}(t,k)+(ik^{2}-ikl'(t))\mu(t,k)=kf(t),\quad 0<t<T,\;\;k\in\mathbb C.
\label{ode1}
\ee
Using this ODE it is shown in Section 2 that if $F(k)$ is defined in 
terms of $f(s)$ by the integral 
\be
F(k)=\int_{0}^{T} \re^{ik^{2}s-ikl(s)}f(s)ds,\quad k\in\mathbb C,
\label{dt1}
\ee
then $f(t)$ can be obtained in terms of $F(k)$ through the solution 
of the following Volterra integral equation
\be
f(t)=\frac{2}{3\pi}\int_{\partial\Omega_{2}^{-}(t)}\re^{-ik^{2}t+ikl(t)}F(k)kdk
     +\frac{2}{3\pi}\int_{0}^{t}J(s,t)f(s)ds,\quad 0<t<T,
\label{genvolt}
\ee
where  $J(s,t)$ is defined by 
equation (\ref{Jdef1}) and $\partial\Omega_{2}^{-}(t)$ is given by
\be
\partial \Omega_{2}^{-}(t)=
\{k\in\mathbb R:\; k\leq 
 l'(t)/2\}\cup\{k=k_{R}+ik_{I}:\;k_{R}=l'(t)/2,\;k_{I}\leq 0\}
\label{omega2-}
\ee
with the orientation shown in Figure \ref{fig2a}.
 
\smallskip
{\bf 3.} {\em Use the global relation (\ref{ex1gr}) and the inversion 
formula (\ref{genvolt}) to derive a Volterra integral equation for 
the unknown boundary values. }
The global relation (\ref{ex1gr}) is of the form (\ref{dt1}) where 
 $f(t)$ and $F(k)$ are replaced respectively by $if_{1}(t)$ and by
 \be
 \hat{q}_{0}(k)-\re^{ik^{2}T}\int_{l(T)}^{\infty}\re^{-ikx}q(x,T)dx+\int_{0}^{T}
 \re^{ik^{2}s-ikl(s)}(k-l'(s))f_{0}(s)ds.
\label{newf} \ee
Furthermore, the global relation is valid for $Im(k)\leq 0$, therefore 
it is valid for $k\in\,\partial\Omega_{2}^{-}(t)$.
It is shown in Section \ref{DtN} that equation (\ref{volterra1}) 
follows from equation (\ref{genvolt}) by replacing in equation 
(\ref{dt1}) $f(s)$ and $F(k)$ by $if_{1}$ and by the expression  in 
 (\ref{newf}).

It turns out that the second term in 
(\ref{newf}) yields a zero contribution, which is consistent  with the 
evolutionary nature of equation (\ref{lsc}) ($q(x,t)$ {\em cannot} 
depend on the future time $T$).

Although the generalised Dirichlet to Neumann map is obtained under the 
assumption of the existence of a unique solution, 
this map can be justified a posteriori {\em without} this assumption,
see Theorem 1.1 of \cite{fpel}.

\section{The spectral analysis of ordinary differential equations and the inversion of complex integrals}
\setcounter{equation}{0}

\begin{prop}
    Let $F(k)$ be defined in terms of $f(t)$ by equation (\ref{dt1}), 
    where $f(t)$ is a sufficiently smooth function. Then $f(t)$ 
    satisfies the Volterra integral equation (\ref{genvolt}). 
    \label{mainprop}
\end{prop}

{\bf Proof:} 
{\bf (a)} We first treat the ODE (\ref{ode1}) as an equation which defines 
$\mu(t,k)$ in terms of $f(t)$ and we seek a solution which is bounded for {\em all} values of the complex 
parameter $k$. 

By integrating with respect to $t$ from either $0$ or 
$T$ we find the following two particular solutions of (\ref{ode1}):
\bea
  \mu_{1}(t,k)&=&\int_{0}^{t}\re^{i(k^{2}(s-t)-k(l(s)-l(t))}kf(s)ds,
  \label{mu1} \\
  \mu_{2}(t,k)&=&-\int_{t}^{T}\re^{i(k^{2}(s-t)-k(l(s)-l(t))}kf(s)ds.
   \label{mu2} \eea
The functions $\mu_{1}$ and $\mu_{2}$ are entire functions of $k$, 
which are bounded, respectively, in the domains $\Omega_{1}$ and $\Omega_{2}$ 
 defined  by
\bea
\Omega_{1}(t)&=&\{k\in\mathbb C:\; k_{I}\geq 0,\; k_{R}\leq\frac{l'(0)}{2}\}\cup 
\{k\in\mathbb C:\; k_{I}<0,\; k_{R}\geq\frac{l'(t)}{2},\;0<t<T\}.
\nonumber \\
\label{omega1}\\
\Omega_{2}(t)&=&\{k\in\mathbb C:\; k_{I}\geq 0,\; k_{R}>\frac{l'(T)}{2}\}\cup 
\{k\in\mathbb C:\; k_{I}\leq 0,\; k_{R}\leq \frac{l'(t)}{2},\; 0<t<T\}.
\nonumber \\
\label{omega2}
\eea

\begin{figure}
    \begin{center} 
 	\includegraphics[width=9cm]{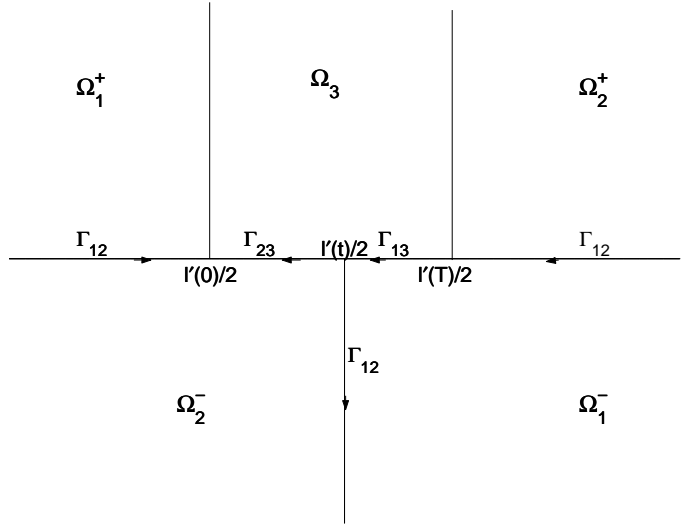} 
    \caption{The domains $\Omega_{1}$, $\Omega_{2}$ and $\Omega_{3}$ 
    in the $k$-plane and the contours $\Gamma_{12}$,$\Gamma_{13}$ and 
    $\Gamma_{23}$}
    \label{fig2b}
    \end{center}
    \end{figure}

The domains $\Omega_{1}(t)$ and $\Omega_{2}(t)$, which are depicted in figure 
\ref{fig2b},  are determined by the real part of the exponent of the 
exponential term appearing in 
equations (\ref{mu1}),(\ref{mu2}), which equals
$$
\re^{-(s-t)k_{I}(2k_{R}-\frac{l(s)-l(t)}{s-t})}=\re^{-(s-t)k_{I}(2k_{R}-l'(\tau))},
$$
where $\tau$ is in the interval bounded by $s$ and $t$. 

For $\mu_{1}$, $s-t\leq 0$, thus $\mu_{1}$ is bounded if and only if
$$
k_{I}(2k_{R}-l'(\tau))\leq 0,
$$
i.e.
$$
\{k_{I}\geq 0,\; k_{R}\leq\frac{l'(\tau)}{2}\} \;{\rm or}\;\{k_{I}\leq 0,\; 
k_{R}\geq \frac{l'(\tau)}{2}\},
$$
for every $\tau$ in the interval $0<s<\tau<t$. Taking into account that $l'(t)$ is an 
increasing function, the above inequalities yield the definition of $\Omega_{1}$. 
Similarly for $\mu_{2}$ and $\Omega_{2}(t)$.

A solution of the ODE (\ref{ode1}) which is bounded in the domain 
\be
\Omega_{3}=\{k\in\mathbb C:\; k_{I}\geq 0,\; 
\frac{l'(0)}{2}\leq k_{R}\leq\frac{l'(T)}{2}\},
\label{omega3}
\ee
is given by
\be
\mu_{3}(t,k,\bar k)=\int_{S(k_{R})}^{t}\re^{i(k^{2}(s-t)-k(l(s)-l(t))}kf(s)ds,
   \label{mu3}
\ee
where the function $S(k_{R})$ is defined on the real interval $[\frac{l'(0)}{2},\frac{l'(T)}{2}]$
by
\be
S(k_{R}):[\frac{l'(0)}{2},\frac{l'(T)}{2}]\to [0,T], \quad S(k_{R})=s \Longleftrightarrow 
k_{R}=\frac{l'(s)}{2}.
\label{Sk}
\ee
In order to prove that the function $\mu_{3}$ is bounded in 
$\Omega_{3}$, we distinguish two cases:
\begin{itemize}
    \item
    $0\leq S(k_{R})<t$
    
    In this case,  $s-t\leq 0$ and since $k_{I}\geq 0$, we need to prove 
    that $k_{R}\leq l'(\tau)/2$, which follows from the fact that 
    $l'(\tau)/2>l'(s)/2\geq k_{R}$.
    \item
    $t<S(k_{R})\leq T$
    
    In this case,  $s-t\geq 0$ and since $k_{I}\geq 0$, we need to prove 
	that $k_{R}\geq l'(\tau)/2$, which follows from the fact that 
	$l'(\tau)/2<l'(s)/2\leq k_{R}$. 
\end{itemize}

We emphasise that the function $\mu_{3}(t,k,\bar k)$, in contrast with the 
functions $\mu_{1}(t,k)$ and $\mu_{2}(t,k)$, involves  $k_{R}$, 
hence this function
dependes on both $k$ and $\bar k$.

Integration by parts of the equations defining the $\mu_{j}$'s 
implies the following asymptotic behaviour
\be\mu_{j}=O\left(\frac{1}{k}\right),\quad k\in \; \Omega_{j},\quad 
k\to\infty,\quad j=1,2,3.
\label{asymp}
\ee
Equations (\ref{mu1}), (\ref{mu2}) and (\ref{mu3}) define a function 
$\mu(t,k,\bar k)$ in terms of $f(t)$.

\smallskip
{\bf (b)} Using the fact that the 
function $\mu(t,k,\bar k)$ is bounded in the entire complex $k$ plane, including  
infinity (see equation (\ref{asymp})), it is possible to 
find an alternative representation for this function using the
Pompeiu ( also known as Cauchy-Green, or d-bar) formula \cite{afok},
\be
\mu(t,k,\bar k)=\frac{1}{2\pi i}\int_{\gamma(t)} \frac{\rm ``jumps''}{\lambda 
-k}d\lambda+\frac{1}{2\pi i}\int\int_{\Omega(t)}\frac{\partial 
\mu(t,\lambda,\bar 
\lambda)}{\partial \bar\lambda}
\frac{d\lambda \wedge d\bar\lambda}{\lambda -k},\quad 
0<t<T,\,k\in\mathbb C,
\label{mugen}
\ee
where $d\lambda\wedge d\bar{\lambda}=-2id\lambda_{R}d\lambda_{I}$,
$\gamma(t)$ denotes the contour along which the function $\mu$ has 
a ``jump'' discontinuity and $\Omega(t)$ is the domain where $\partial 
\mu/\partial \bar k\neq 0$. Before computing the relevant jumps, we note 
that $\mu_{3}$ coincide with $\mu_{1}$ on the half-line 
$\{k_{I}>0,\;k_{R}=l'(0)/2\}$  and with $\mu_{2}$ on the half-line $\{k_{I}>0,\;k_{R}=l'(T)/2\}$.
This is consequence of the definition of $S(k_{R})$, which implies 
that $S(l'(s)/2)=s$ for all $0\leq s\leq T$.
Thus, the relevant jumps are given by the expressions
\bea
\mu_{1}-\mu_{2}&=&\int_{0}^{T}A(k,s,t,)ds,\quad k\in\,\Gamma_{12},
  \nonumber \\
  \mu_{1}-\mu_{3}&=& \int_{0}^{S(k_{R})}A(k,s,t,)ds,  \quad 
  k\in\,\Gamma_{13}, \nonumber \\
  \mu_{2}-\mu_{3}&=& \int_{T}^{S(k_{R})}A(k,s,t,)ds,\quad 
  k\in\,\Gamma_{23},
    \label{jumps} \eea
where we use the following notation (see figure \ref{fig2b}):
\bea
A(k,s,t)&=&\re^{ik^{2}(s-t)-ik(l(s)-l(t))}kf(s),
\label{Adef} \\
\Gamma_{12}(t)&=&\{k_{I}=0, 
\;k_{R}\in\,(-\infty,\frac{l'(0)}{2})\}\cup\{k_{R}=\frac{l'(t)}{2},\;k_{I}\in\,(0,-\infty)\}\cup
\nonumber \\
&&\cup\{
k_{I}=0, \;k_{R}\in (\infty, l'(T)/2)\},
\nonumber\\
\Gamma_{13}(t)&=&\{k_{I}=0, 
\;k_{R}\in\,(\frac{l'(T)}{2},\frac{l'(t)}{2})\},
\label{gammas}\\
\Gamma_{23}(t)&=&\{k_{I}=0, 
\;k_{R}\in\,(\frac{l'(t)}{2},\frac{l'(0)}{2})\}.
\nonumber
\eea
The direction of integration is depicted in figure \ref{fig2b}, and it 
is also indicated in equations (\ref{gammas});  for example, 
$k_{R}\in(-\infty,l'(0)/2)$ indicates that the integration is from 
$-\infty$ to \, $l'(0)/2$.

Equation (\ref{mu3}) implies 
\be
\frac{\partial \mu_{3}(t,k,\bar k)}{\partial \bar k}=-\frac{\partial
S(k_{R})}{\partial \bar 
k}k\{\re^{ik^{2}(s-t)-k(l(s)-l(t))}f(s)\}_{s=S(k_{R})}=-\frac{\partial 
S(k_{R})}{\partial \bar k}A(k,S(k_{R}),t).
    \label{dbarder}
    \ee
Hence, using equations (\ref{jumps}) and (\ref{dbarder}) in 
equation (\ref{mugen}), we find the following alternative 
representation of $\mu$:
\bea
\mu(t,k,\bar k)=\frac{1}{2\pi 
i}\int_{\Gamma_{12}}\left(\int_{0}^{T}A(\lambda,s,t)ds\right) \frac{d\lambda}{\lambda 
-k}+\frac{1}{2\pi 
i}\int_{\Gamma_{13}}\left(\int_{0}^{S(k_{R})}A(\lambda,s,t)ds\right) \frac{d\lambda}{\lambda 
-k}
\nonumber \\
+\frac{1}{2\pi 
i}\int_{\Gamma_{23}}\left(\int_{T}^{S(k_{R})}A(\lambda,s,t)ds\right) \frac{d\lambda}{\lambda 
-k}-\frac{1}{2\pi i}\int\int_{\Omega_{3}}\frac{\partial S(\lambda_{R})}{\partial \bar\lambda}
A(\lambda,S(\lambda_{R}),t)\frac{d\lambda \wedge d\bar\lambda}{\lambda -k}.
\label{mu}
\eea

\smallskip
{\bf (c)} The representation of the function $\mu(t,k,\bar k)$  
defined by equations (\ref{mu1}), (\ref{mu2}), (\ref{mu3}) involves 
$f(t)$, while the represnetation defined by equation (\ref{mu}) 
involves various integrals of $A(k,s,t)$. Thus there exists a relation 
between $f(t)$ and these integrals of $A$. The simplest way to obtain 
this relation is to consider the large $k$ asymptotic behaviour of 
$\mu(t,k,\bar k)$. Equation (\ref{mu}) implies
$$
\mu(t,k,\bar k)=\frac{\mu_{0}(t)}{k}+O\left(\frac{1}{k}\right),\quad 
\mu_{0}=\lim_{k\to\infty} (k\mu).
$$
Substituting this expression in the ODE 
(\ref{ode1}) we find $\mu_{0}=f(t)$, thus
\bea
f(t)=\frac{1}{2\pi}\left\{-\int_{\Gamma_{12}}\left(\int_{0}^{T}A(k,s,t)ds \right)dk- 
\int_{\Gamma_{13}}\left(\int_{0}^{S(k_{R})}A(k,s,t)ds\right)dk
\right.\nonumber \\
-\left.\int_{\Gamma_{23}}\left(\int_{T}^{S(k_{R})}A(k,s,t)ds\right) dk
+\int\int_{\Omega_{3}}\frac{\partial S(k_{R})}{\partial 
\bar k}
A(k,S(k_{R}),t)dk \wedge d\bar k\right\}.
\label{fform}
\eea
Equation (\ref{fform}) was first derived in \cite{fpel} (see equation (3.2) 
of \cite{fpel}). 

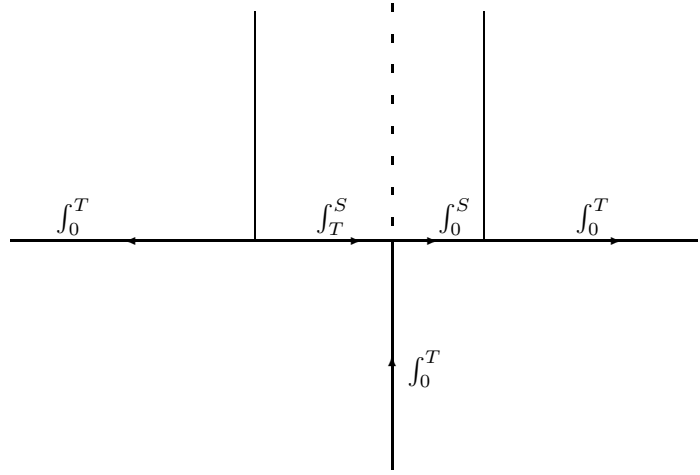
\begin{figure}[h]
    \begin{center}
	\setlength{\unitlength}{0.04in}
	\begin{picture}(90,30)(-50,6)
      \thinlines
   \put(-20,35){\vector(-1,0){25}} 
   \put(-20,35){\vector(1,0){6}}
   \put(-14,35){\line(1,0){4}}
   \put(-10,35){\vector(1,0){6}}
   \put(-4,35){\line(1,0){4}}
   \put(0,35){\vector(1,0){20}}
   \put(20,35){\line(1,0){10}}
    \put(-35,35){\line(-1,0){25}}
    \put(-10,35){\line(0,-1){30}}
    \put(-28,35){\line(0,1){30}}
    \put(2,35){\line(0,1){30}}
    \put(-10,15){\vector(0,1){5}}
    \put(-10,37){\line(0,1){1}}
    \put(-10,41){\line(0,1){1}}
    \put(-10,45){\line(0,1){1}}
    \put(-10,49){\line(0,1){1}}
    \put(-10,53){\line(0,1){1}}
    \put(-10,57){\line(0,1){1}}
    \put(-10,61){\line(0,1){1}}
    \put(-10,65){\line(0,1){1}}
    \put(-4,37){$\int_{0}^{S}$}
    \put(-20,37){$\int_{T}^{S}$}
    \put(14,37){$\int_{0}^{T}$}
    \put(-8,17){$\int_{0}^{T}$}
    \put(-54,37){$\int_{0}^{T}$}
     \end{picture}
     \caption{The integrands in equation (\ref{fform}) }
% \hspace{20pt}(2b) A 
%     curve $\Gamma(t)$}
 \label{fig3} 
 \end{center}
     \end{figure}

\begin{figure}[t]
	 \begin{center}
	     \setlength{\unitlength}{0.04in}
	     \begin{picture}(90,65)(-50,10)
	   \thinlines
	\put(-20,35){\line(-1,0){25}} 
	\put(-20,35){\line(1,0){6}}
	\put(-14,35){\line(1,0){4}}
	\put(-10,35){\vector(1,0){6}}
	\put(-4,35){\line(1,0){4}}
	\put(0,35){\line(1,0){20}}
	\put(20,35){\line(1,0){10}}
	 \put(-35,35){\line(-1,0){25}}
	 \put(-10,35){\line(0,-1){30}}
	 \put(2,35){\vector(0,1){20}}
	 \put(2,55){\line(0,1){10}}
	 \put(-10,15){\line(0,1){5}}
	 \put(-28,35){\line(0,1){30}}
	 \put(4,48){$\int_{T}^{t}$}
	 \put(-4,37){$\int_{S}^{t}$}
	 \put(-10,37){\line(0,1){1}}
	     \put(-10,41){\line(0,1){1}}
	     \put(-10,45){\line(0,1){1}}
	     \put(-10,49){\line(0,1){1}}
	     \put(-10,53){\line(0,1){1}}
	     \put(-10,57){\line(0,1){1}}
	     \put(-10,61){\line(0,1){1}}
	     \put(-10,65){\line(0,1){1}}
	  \end{picture}
	  \caption{The integrands in equation (\ref{green2}) }
     % \hspace{20pt}(2b) A 
     %     curve $\Gamma(t)$}
      \label{fig4} 
      \end{center}
	  \end{figure}
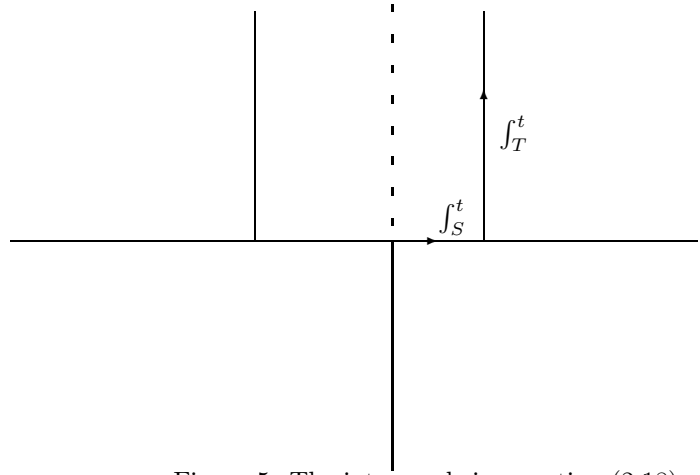

\smallskip
{\bf (d)} We will now show that equation (\ref{fform}) can be transformed into 
a Volterra integral equation. For this purpose, we split 
the domain $\Omega_{3}$ in the form $\Omega_{3}=\Omega_{3}^{(1)}(t)\cup 
\Omega_{3}^{(2)}(t)$ where
\be
\Omega_{3}^{(1)}(t)=\{k\in\Omega_{3}:l'(0)/2<k_{R}<l'(t)/2\},\quad
\Omega_{3}^{(2)}(t)=\{k\in\Omega_{3}:l'(t)/2<k_{R}<l'(T)/2\}.
\label{omega3split}
\ee
In the domain $\Omega_{3}^{(2)}(t)$ we use the complex form of Green's theorem, 
which states that
\be
	  -\int\int_{\Omega_{3}^{(2)}(t)}\frac{\partial \mu_{3}(k,\bar 
	  k)}{\partial \bar k}
	  dk \wedge d\bar k=\int_{\partial 
	  \Omega_{3}^{(2)}(t)}\mu_{3}(k, \bar{k})dk,
	  \label{green}
	  \ee
where the boundary $\partial \Omega_{3}^{(2)}(t)$ of 
$\Omega_{3}^{(2)}(t)$
has counterclockwise orientation.
Recalling that $S(k_{R})=t$ when $2k_{R}=l'(t)$, it follows that the 
contribution of the half line $\{k_{R}=l'(t)/2, k_{I}>0\}$ to the 
right hand side of equation (\ref{green}) vanishes, hence
$$
\int\int_{\Omega_{3}^{(2)}(t)}\frac{\partial S(k_{R})}{\partial 
\bar k}
A(k,S(k_{R}),t)dk \wedge d\bar k=-\int\int_{\Omega_{3}^{(2)}(t)}
\frac{\partial \mu_{3}(k,\bar 
	  k)}{\partial \bar k}
	  dk \wedge d\bar k=
$$
\be
=\int_{l'(t)/2}^{l'(T)/2}\int_{S(k_{R})}^{t}A(k,s,t)ds 
dk+\int_{l(T)}\int_{T}^{t}A(k,s,t)ds dk,
\label{green2}
\ee
where the half line $l(T)$ is defined by 
\be
l(T)=\{k:k_{R}=l'(T)/2,\;0\leq k_{I}<\infty\}.
\label{gamma-T}
\ee
The integrands of the line integrals appearing on the right hand side 
of equation (\ref{fform}) are shown in figure \ref{fig3}, while the 
integrands  of the line integrals appearing on the right hand side 
of equation (\ref{green2}) are shown in figure \ref{fig4}.

\begin{figure}[t]
    \begin{center}
	\setlength{\unitlength}{0.04in}
	\begin{picture}(90,20)(-50,5)
      \thinlines
   \put(-20,35){\vector(-1,0){25}} 
   \put(-20,35){\line(1,0){6}}
   \put(-10,35){\vector(-1,0){4}}
   \put(-10,35){\vector(1,0){6}}
   \put(-4,35){\line(1,0){4}}
   \put(0,35){\vector(1,0){20}}
   \put(20,35){\line(1,0){10}}
   \put(2,34){\vector(1,0){25}}
      \put(25,34){\line(1,0){5}}
    \put(-35,35){\line(-1,0){25}}
    \put(-10,35){\line(0,-1){30}}
    \put(2,65){\vector(0,-1){20}}
    \put(2,45){\line(0,-1){10}}
    \put(-10,15){\vector(0,1){5}}
    \put(-28,35){\line(0,1){30}}
    \put(4,48){$\int_{t}^{T}$}
    \put(-28,36){\vector(1,0){14}}
    \put(-14,36){\line(1,0){4}}
    \put(-20,38){$\int_{0}^{S}$}
    \put(-5,37){$\int_{0}^{t}$}
    \put(-20,30){$\int_{0}^{T}$}
       \put(14,37){$\int_{t}^{T}$}
       \put(24,30){$\int_{0}^{t}$}
       \put(-8,17){$\int_{0}^{T}$}
       \put(-54,37){$\int_{0}^{T}$}
    \put(-10,37){\line(0,1){1}}
	\put(-10,41){\line(0,1){1}}
	\put(-10,45){\line(0,1){1}}
	\put(-10,49){\line(0,1){1}}
	\put(-10,53){\line(0,1){1}}
	\put(-10,57){\line(0,1){1}}
	\put(-10,61){\line(0,1){1}}
	\put(-10,65){\line(0,1){1}}
     \end{picture}
     \caption{The integrands in the expression obtained by adding 
     (\ref{fform}) and (\ref{green2}) together. The double lines 
     indicate that the contribution is split into two integrals.
     }
% \hspace{20pt}(2b) A 
%     curve $\Gamma(t)$}
 \label{fig5} 
 \end{center}
     \end{figure}
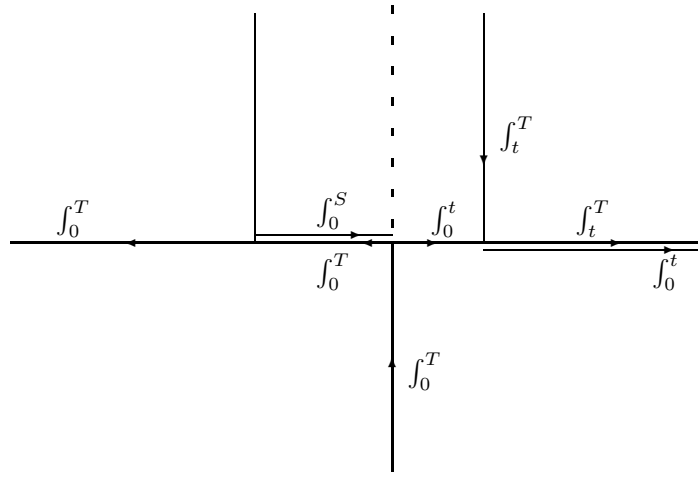

Adding the integrands in the respective line integrals, we find the 
contributions depicted in figure (\ref{fig5}).

The above analysis, together with the identity
$$
\frac{\partial S(k_{R})}{\partial 
\bar k}=\frac{d S(k_{R})}{2
d k_{R}},\quad dk\wedge d\bar k=-2i dk_{R}dk_{I}.
$$
implies that equation (\ref{fform}) can be rewritten in the following 
form
\bea
2\pi f(t)&=&\int_{\partial 
\Omega_{2}^{-}(t)}\left(\int_{0}^{T}A(k,s,t)ds\right) 
\,dk
+\int_{\partial \Omega_{2}^{+}(t)}\left(\int_{t}^{T}A(k,s,t)ds\right)dk
\nonumber \\
&+&\int_{l'(0)/2}^{l'(t)/2}\left(\int_{0}^{S(k_{R})}A(k,s,t)ds\right) dk
+\left\{\int_{l'(t)/2}^{\infty}\left(\int_{0}^{t}A(k,s,t)ds\right) dk
\right.\nonumber \\
&-&i\left.
\int_{0}^{\infty}\left(\int_{l'(0)/2}^{l'(t)/2}\left(\frac{dS(k_{R})}{dk_{R}}\right)
A(k,S(k_{R}),t)dk_{R}\right)dk_{I},\right\}.
\label{fformfinb}
\eea
As a consequence of splitting the integrals originally appearing together on the 
right hand side of equation (\ref{mu}), the resulting individual 
integrals in equation (\ref{fformfinb}) have a singularity as 
$k\to\infty$. Hence the last two integrals  on the right hand 
side of equation (\ref{fformfinb}) must be considered together (this 
is indicated by the curly brackets around these terms).

The first term on the right hand side of this expression can be 
computed in terms of the given data $F(k)$, defined in (\ref{dt1}), 
hence this term is known. 

The second term equals $(\pi/2)f(t)$. Indeed, 
the integrand in this term is bounded and analytic in 
$\Omega_{2}^{+}(t)$, and it behaves like $if(t)/k$ as  $k\to\infty$. 
Thus this term equals
$$
if(t)\int_{\pi/2}^{0}id\theta=\frac{\pi}{2}f(t).
$$

The third integral in (\ref{fformfinb}) can be written in  a ``Volterra 
form''  by exchanging the order of integration. 
Indeed, recalling the definition of $S(k_{R})$ and changing variable 
to $\sigma=S(k_{R})$ so that $k(=k_{R})=l'(\sigma)/2$, 
we find that this integral is equal 
to 
$$
\int_{l'(0)/2}^{l'(t)/2}\left(\int_{0}^{S(k_{R})}A(k,s,t)ds\right) dk
=
\int_{0}^{t}\left(\int_{l'(s)/2}^{l'(t)/2}A(k,s,t) 
dk\right)ds
$$
In the double 
integral we change the variable $k_{R}$ to $s=S(k_{R})$ and rename 
$k_{I}$ as $k$. This yields
$$
\int_{0}^{\infty}\int_{l'(0)/2}^{l'(t)/2}\frac{dS(k_{R})}{dk_{R}}
A(k,S(k_{R}),t)dk_{R}dk_{I}=\int_{0}^{\infty}\left(
\int_{0}^{t}A(\frac{l'(s)}{2}+ik, s,t)ds \right)dk.
$$
Hence equation (\ref{fformfinb}) yields
\be
\frac{3\pi}{2}f(t)=\int_{\partial 
\Omega_{2}^{-}(t)}\left(\int_{0}^{T}A(k,s,t)ds \right)dk+\int_{0}^{t}\left
(\int_{l'(s)/2}^{l'(t)/2}A(k,s,t) 
dk\right)ds+
\label{than1} \ee
$$
\left\{\int_{l'(t)/2}^{\infty}\left(\int_{0}^{t}A(k,s,t)ds\right) dk
-i\int_{0}^{\infty}\left(
\int_{0}^{t}A(\frac{l'(s)}{2}+ik, s,t)ds \right)dk\right\}.
$$
Setting
\be
A(k,s,t)=E(k,s,t)kf(s),\quad
E(k,s,t)=\re^{ik^{2}(s-t)-ik(l(s)-l(t))},
\label{Edef}\ee
and using 
$$
E(\frac{l'(s)}{2}+ik, s,t)=E(ik,s,t)\re^{-k(s-t)l'(s)}G(s,t),
$$
where $G(s,t)$ is given by (\ref{Gst}), equation (\ref{than1}) becomes
\be
\frac{3\pi}{2}f(t)=\int_{\partial 
\Omega_{2}^{-}(t)}\left(\int_{0}^{T}A(k,s,t)ds \right)dk+\int_{0}^{t}
\left(\int_{l'(s)/2}^{l'(t)/2}E(k,s,t) k
dk\right)f(s)ds+
\label{than2} \ee
$$
\left\{\int_{l'(t)/2}^{\infty}k\left(\int_{0}^{t}E(k,s,t)f(s)ds\right) dk
-i\int_{0}^{\infty}\left(\int_{0}^{t}
E(ik,s,t)(k-i\frac{l'(s)}{2})G(s,t)\re^{-k(s-t)l'(s)}f(s)ds\right)dk\right\}.
$$
In what follows we will rewrite the terms in the above bracket in a 
Volterra form. In this regard we note that this bracket is well defined. 
Indeed, 
 integration by parts of the inner integrals yields
$$
\int_{l'(t)/2}^{\infty}\left[\frac{f(t)}{ik}+O\left(\frac{1}{k^{2}}\right)\right]dk
\nonumber \\
-i
\int_{0}^{\infty}
\left[-\frac{f(t)}{k}+O\left(\frac{1}{k^{2}}\right)\right] dk,
$$
which shows that the contribution of the singular term $1/k$ at 
infinity cancels out. 

We start by rewriting the bracket as
\be
\int_{0}^{t}\left[J_{1}(s,t)+G(s,t)J_2(s,t)-
\frac{i}{2}G(s,t)l'(s)\tilde{J}_{2}(s,t)\right]f(s)ds
\label{than3}
\ee
where
\bea
J_{1}(s,t)&=&\lim_{\varepsilon \to 
0^{+}}\int_{l'(t)/2}^{\infty}\re^{ik^{2}(s-t+i\varepsilon)-ik(l(s)-l(t))}k dk,\nonumber\\
J_{2}(s,t)&=&
\lim_{\varepsilon \to 
0^{+}}\int_{0}^{\infty}\re^{-ik^{2}(s-t+i\varepsilon)+k(l(s)-l(t))-k(s-t)l'(s)}
k dk,
\nonumber \\
\tilde{J}_{2}(s,t)&=&\int_{0}^{\infty}E(ik,s,t)\re^{-k(s-t)l'(s)}dk.
\label{Jsdef}
\eea
The kernel in the expression (\ref{than3}) equals
\be
\frac{H(s,t)}{2i(s-t)}+\frac{l(s)-l(t)}{2(s-t)}\left[\tilde 
J_{1}(s,t)-iG(s,t)\tilde J_{2}(s,t)\right]
\label{than4}\ee
where $G(s,t)$ is given by (\ref{Gst}) and
\be
H(s,t)=G(s,t)-\re^{i\frac{l'(t)^2}{4}(s-t)-i\frac{l'(t)}{2}(l(s)-l(t))},\quad
% \nonumber \\
\tilde J_{1}(s,t)=\int_{l'(t)/2}^{\infty}E(k,s,t)dk.
\label{than5}
\ee
Indeed, integrating by parts the expressions for $J_{1}$ and $J_2$ we 
find
\bea
J_{1}(s,t)&=&-\frac{\re^{i\frac{l'(t)^2}{4}(s-t)-i\frac{l'(t)}{2}(l(s)-l(t))}}{2i(s-t+i0)}+\frac{l(s)-l(t)}{2(s-t)}
\int_{l'(t)/2}^{\infty}E(k,s,t)dk,
\label{J1} \\
J_{2}(s,t)&=&
% \frac{1}{2i(s-t+i0)}+\left[\frac{l(s)-l(t)}{2i(s-t)}-\frac{l'(s)}{2i}\right]
% \int_{0}^{\infty}E(ik,s,t)\re^{-k(s-t)l'(s)}dk=
% \nonumber \\
% &=&
\frac{1}{2i(s-t+i0)}+\left[\frac{l(s)-l(t)}{2i(s-t)}-\frac{l'(s)}{2i}\right]\tilde{J}_{2}(s,t).
\label{J2}
\eea
Substituting the above expressions in (\ref{than3}) we find that the 
kernel equals the expression in (\ref{than4}).

Using integration by parts of the second term on the right hand side 
of (\ref{than2}), we find
\be
\int_{l'(s)/2}^{l'(t)/2}E(k,s,t)kdk=-\frac{H(s,t)}{2i(s-t)}+\frac{l(s)-l(t)}{2(s-t)}
\int_{l'(s)/2}^{l'(t)/2}E(k,s,t)dk.
\label{than6}
\ee
Adding the expressions in (\ref{than4}) and (\ref{than6}) we find that 
the equation (\ref{than2}) can be written in the form
$$
\frac{3\pi}{2}f(t)=\int_{\partial 
\Omega_{2}^{-}(t)}\left(\int_{0}^{T}A(k,s,t)ds 
\right)dk+\int_{0}^{t}J(s,t)f(s)ds,
$$
where 
%%HERE
%%
%%
%%
% \be
% J(s,t)=\frac{l(s)-l(t)}{2(s-t)}\left[\int_{l'(s)/2}^{\infty} 
% E(k,s,t)dk-iG(s,t)\int_{0}^{\infty}E(ik,s,t)e^{-k(s-t)l'(s)}dk\right].
% \label{than7}
% \ee
% This is the kernel given by (\ref{Jdef1}). 
$J(s,t)$ is the kernel given by (\ref{Jdef1}).

\begin{rem}
    {\em 
  The direct  evaluation of  the first integral appearing in expression 
  (\ref{Jdef1}), by completing the square in the exponent, yields
  $$
  \int_{l'(s)/2}^{\infty} 
  E(k,s,t)dk=\re^{i\frac{(l(t)-l(s))^{2}}{4(t-s)}}\int_{l'(s)/2}^{\infty}\re^{-i(t-s)
  [k -\frac{l(s)-l(t)}{2(s-t)}]^{2}}dk$$
  Setting $\lambda=[k-\frac{l(s)-l(t)}{2(s-t)}]\sqrt{t-s}$, and 
$\lambda_{0}(s,t)=\sqrt{t-s}\left[l'(s)/2-\frac{l(s)-l(t)}{2(s-t)}\right]$,
  this 
  can be written as 
  \be
  \int_{l'(s)/2}^{\infty} 
   E(k,s,t)dk=\frac{ \re^{i\frac{(l(t)-l(t))^{2}}{4(t-s)}}}{\sqrt{t-s}}
\int_{\lambda_{0}(s,t)}^{\infty}\re^{-i\lambda^{2}}d\lambda.
   \label{directev}
   \ee
The integral on the right hand side of equation 
(\ref{directev}) can be expressed in terms of the Gamma function, 
and of the imaginary error function Erfi$(z)$. Indeed, 
% this integral is equal to 
$$
\int_{\lambda_{0}(s,t)}^{\infty}\re^{-i\lambda^{2}}d\lambda=\int_{\lambda_{0}(s,t)}^{0}\re^{-i\lambda^{2}}d\lambda+
\int_{0}^{\infty}\re^{-i\lambda^{2}}d\lambda=
$$
$$
=\frac{i+1}{2}\sqrt{\frac{\pi}{2}}{\rm Erfi}(\re^{3\pi 
i/4}\lambda_{0}(s,t))+\frac{\sqrt{\pi}}{2}\re^{-i\pi/4}.
% \Gamma\left(\frac{1}{2}\right).
$$

}
\end{rem}

\section{The Dirichlet to Neumann map for the linear Schr\"odinger 
equation}\label{DtN}
\setcounter{equation}{0}

\begin{prop}
Let the complex-valued scalar function $q(x,t)$ satusfy the following 
initial-boundary value problem:
\be
\begin{array}{lll}
iq_{t}+q_{xx}=0, & l'(t)<x<\infty, &0<t<T,
\\ q(x,0)=q_{0}(x), && l'(t)<x<\infty,
\\ q(l(t),t)=f_{0}(t),&& 0<t<T,
\\ q_{0}(0)=f_{0}(0)&&
\end{array}
\label{dirbvp1}
\ee
where $
l(t)\in{\bf C}^{2}[0,T], \quad l''(t)>0,\quad l(0)=0.
$
The Dirichlet to Neumann map for this problem is characterised by the  linear 
Volterra integral equation (\ref{volterra1}).

\end{prop}

{\bf Proof} In order to obtain the linear integral equation satisfied 
by $f_{1}(t)=q_{x}(l(t),t)$ we must replace in equation (\ref{genvolt})
the function $f(t)$  
 by $if_{1}(t)$, and the function $F(k)$ by the expression in 
 (\ref{newf}). Multiplying the latter expression by $k$ we find three 
 terms. 
 
 The first term in  (\ref{newf}), multiplied by $k$, equals 
 \be
 k\int_{0}^{\infty}\re^{-ikx}q_{0}(x)dx=-iq_{0}(0)-i\int_{0}^{\infty}\re^{-ikx}
 q_{0}'(x)dx.
 \label{firstterm}
 \ee
 
 The second term multiplied by $k\re^{-ik^{2}t+ikl(t)}$ is 
%  an integral along $\partial 
%  \Omega_{2}^{-}(t)$ whose integrand is
\be
k\re^{ik^{2}(T-t)-ik(l(T)-l(t))}\int_{l(T)}^{\infty}\re^{-ik(x-l(T))}q(x,T)dx.
\label{vanint}
\ee
The exponential multiplying the above integral is bounded in the domain 
$\Omega_{2}^{-}(t)$, while the integral in (\ref{vanint}) 
is bounded and analytic for Im$(k)\leq 0$ and is of order 
$O\left(\frac{1}{k}\right)$ as $k\to\infty$. Thus, the application of 
Jordan's lemma (after a suitable change of variables) implies that the 
integral of (\ref{vanint}) along $\partial \Omega_{2}^{-}(t)$ 
vanishes.  

The third term in  (\ref{newf}), multiplied by $k$, equals
\be
\int_{0}^{T}\re^{ik^{2}s-ikl(s)}(k^{2}-kl'(s))f_{0}(s)ds=-i\re^{ik^{2}T-ikl(T)}f_{0}(T)
+if_{0}(0)+i\int_{0}^{T}\re^{ik^{2}s-ikl(s)}f_{0}'(s)ds.
\label{thirdterm}
\ee

Adding the terms on the right hand side of equations 
(\ref{firstterm}) and (\ref{thirdterm}) (using the compatibility 
$q_{0}(0)=f_{0}(0)$), multiplying the resulting 
expression by $\re^{-ik^{2}t+ikl(t)}$ and integrating it with respect 
to $k$ along the contour $\partial 
 \Omega_{2}^{-}(t)$, we find
$$
 f_{1}(t)=\frac{2}{3\pi}\int_{\partial 
 \Omega_{2}^{-}(t)}\re^{-ik^{2}t+ikl(t)}\left[\int_{0}^{t}\re^{ik^{2}s-ikl(s)}
 f_{0}'(s)ds
 -\int_{0}^{\infty}\re^{-ikx}q_{0}'(x)dx\right]dk
 $$
 \be
 +\frac{2}{3\pi}\int_{0}^{t}J(s,t)f_{1}(s)ds,\quad 0<t<T.
 \label{finvolt1}
  \ee
 Indeed, the integral involving $f_{0}(T)$ as well as the integral 
 involving $\int_{t}^{T}\re^{ik^{2}s-ikl(s)}f_{0}'(s)ds$ vanish, 
 because the term $\exp[-ik^{2}(t-s)+ik(l(t)-l(s))]$ is bounded and 
 analytic in $\Omega_{2}^{-}(t)$ whenever $s-t\geq 0$. 
 
\begin{figure}
     \begin{center}
	 \setlength{\unitlength}{0.04in}
	 \begin{picture}(90,30)(-50,6)
      \thinlines
    \put(-20,35){\vector(-1,0){25}} 
    \put(-45,35){\line(-1,0){25}}
    \put(-20,35){\line(0,-1){30}}
    \put(-20,15){\vector(0,1){5}}
    \put(-22,37){$a(t,s)$}
    \put(-30,41){\line(0,1){1}}
    \put(-30,45){\line(0,1){1}}
    \put(-30,49){\line(0,1){1}}
    \put(-30,53){\line(0,1){1}}
    \put(-30,57){\line(0,1){1}}
    \put(-30,37){\line(0,1){1}}
	\put(-30,33){\line(0,1){1}}
	\put(-30,29){\line(0,1){1}}
	\put(-30,25){\line(0,1){1}}
	\put(-30,21){\line(0,1){1}}
	\put(-30,17){\line(0,1){1}}
		\put(-30,13){\line(0,1){1}}
		\put(-30,9){\line(0,1){1}}
		\put(-30,5){\line(0,1){1}}
    \put(40,35){\vector(-1,0){25}} 
    \put(15,35){\line(-1,0){15}}
    \put(40,35){\line(0,-1){30}}
    \put(40,15){\vector(0,1){5}}
    \put(42,37){$b(t,x)$}
    \put(20,41){\line(0,1){1}}
    \put(20,45){\line(0,1){1}}
    \put(20,49){\line(0,1){1}}
    \put(20,53){\line(0,1){1}}
    \put(20,57){\line(0,1){1}}
    \put(20,37){\line(0,1){1}}
	\put(20,33){\line(0,1){1}}
	\put(20,29){\line(0,1){1}}
	\put(20,25){\line(0,1){1}}
	\put(20,21){\line(0,1){1}}
	\put(20,17){\line(0,1){1}}
		\put(20,13){\line(0,1){1}}
		\put(20,9){\line(0,1){1}}
		\put(20,5){\line(0,1){1}}
     \end{picture}
     \caption{(7a) The domain $\tilde\Omega_{2}^{-}(t,s)$ 
 %     in the $k$ plane 
 \hspace{20pt}(7b) The domain $\tilde\Omega_{2}^{-}(t,x)$}
  \label{fig6} 
 \end{center}
     \end{figure}
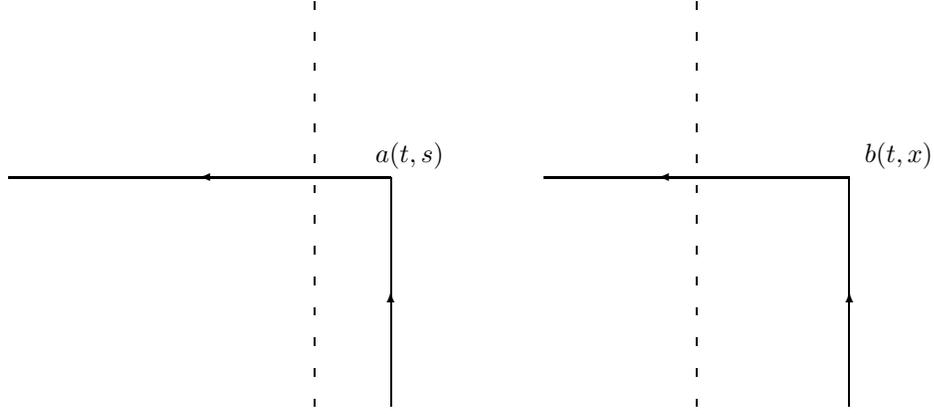

 After exchanging order of integration, the integral over $k$ on the right hand 
 side of equation 
 (\ref{finvolt1}) can be computed explicitly, and when this is done
 equation (\ref{finvolt1}) becomes 
 equation (\ref{volterra1}). 
 Indeed,
 \be
 \frac{2}{3\pi}\int_{\partial 
 \Omega_{2}^{-}(t)}\re^{ik^{2}(s-t)-ik(l(s)-l(t))}dk=\left(\frac{2}{3\pi}
 \int_{\partial 
 \tilde\Omega_{2}^{-}(t,s)}\re^{-i\lambda^{2}}d\lambda\right)
 \frac{\re^{i\frac{(l(t)-l(s))^{2}}{4(t-s)}}}{\sqrt{t-s}}
 \label{def1}
 \ee
 and
 \be
 \frac{2}{3\pi}\int_{\partial 
\Omega_{2}^{-}(t)}\re^{-ik^{2}t+ikl(t)-ikx}dk=\left(\frac{2}{3\pi}
 \int_{\partial 
 \hat\Omega_{2}^{-}(t,x)}\re^{-i\lambda^{2}}d\lambda\right)
 \frac{\re^{i\frac{(l(t)-x)^{2}}{4t}}}{\sqrt{t}},
 \label{def2}
  \ee
 where the domains $\tilde\Omega_{2}^{-}(t,s)$ and 
 $\tilde\Omega_{2}^{-}(t,x)$ are depicted in figure (\ref{fig6}) and 
 the functions $a(t,s)$ and $b(t,x)$ appearing in the figures are given by 
 $$a(t,s)=\frac{\sqrt{t-s}}{2}\left[l'(t)-\frac{l(t)-l(s)}{t-s}\right],\quad 
 b(t,x)=\frac{\sqrt{t}}{2}\left[l'(t)-\frac{l(t)-x}{t}\right].
 $$

 Since $t\geq s$ it follows that $\frac{l(t)-l(s)}{t-s}=l'(\tau)$, 
 $s\leq \tau\leq t$, thus $l'(t)-l'(\tau)\geq 0$ and hence 
 $a(t,s)\geq 0$.
 Similarly, 
 $$
 l'(t)-\frac{l(t)-x}{t}=l'(t)-l'(\tau)+\frac{x}{2},\quad \tau \leq t,
 $$
 thus $b(t,x)$ is also positive.
 
 The exponential $\re^{-i\lambda^{2}}$ is bounded in the 
 fourth quadrant of the complex $\lambda$ plane, so that both the 
 lines $a+ik_{I}$ and $b+ik_{I}$, $k_{I}\leq 0$, can be deformed to 
 the real axis. Hence both the $\lambda$ integrals equal 
 $$
 \frac{2}{3\pi}\int_{-\infty}^{\infty} \re^{-i\lambda^{2}}d\lambda 
 =\frac{2}{3\pi}\left[2\int_{0}^{\infty} 
 \re^{-i\lambda^{2}}d\lambda\right]= 
 \frac{2}{3\pi}\re^{-i\pi/4}\Gamma\left(\frac{1}{2}\right)= 
 \frac{2}{3\sqrt{\pi}}\re^{-i\pi/4}.
$$

{\bf QED}
%  $$
%  \tilde\Omega_{2}^{-}(t,s)=\{\tilde k_{R}+i\tilde k_{I}:\;\; \tilde 
%  k_{R}\leq 
%  \left(\frac{l'(t)}{2}-\frac{l(t)-l(s)}{2(t-s)}\right)\sqrt{t-s},\;\tilde 
%  k_{I}\leq 0\},
%  $$
%  $$
%  \tilde\Omega_{2}^{-}(t,x)=\{\tilde k_{R}+i\tilde k_{I}:\;\; \tilde 
%   k_{R}\leq 
%   \left(\frac{l'(t)}{2}-\frac{l(t)-x}{2t}\right)\sqrt{t},\;\tilde k_{I}\leq 0\}.
%   $$
% The bounds 
%   for $\tilde k_{R}$ can be written in terms of some $\tau\leq t$ 
%   (not the same in the two expressions) as follows 
%   $$
%   \frac{l'(t)}{2}-\frac{l(t)-l(s)}{2(t-s)}=\frac{l'(t)-l'(\tau)}{2}, 
%  \quad \quad
%   \frac{l'(t)}{2}-\frac{l(t)-x}{2t}=\frac{l'(t)-l'(\tau)}{2}+\frac{x}{2t}.
%  $$
%  Since $l'(t)\geq l'(\tau)$ and both $x$ and $t$ are positive, and 
%  the exponential $\re^{-i\lambda^{2}}$ is analytic and bounded 
%   in the fourth quadrant of the $\lambda$ complex plane, we can 
%  use analyticity to deform both contours of integration in 
%  (\ref{def1}) and (\ref{def2}) to the time-independent contour 
%  (\ref{tq}). 

 \section{Conclusions}
\setcounter{equation}{0}
The results presented in this paper are perhaps interesting in a 
broader context, as they illustrate the implementation of a new
technique, which allows one to invert complicated integrals, such as 
the integral defined by equation (\ref{dt1}). This integral is  
a simple variant of the elementary integral
\be
{\cal F}(k)=\int_{0}^{T}\re^{ik^{2}s}f(s)ds, \quad k\in\mathbb C.
\label{ft1}
\ee
This integral can be inverted by a straightforward application of the 
inverse  
Fourier transform (after a suitable change of variables)
\be
f(t)=\frac{1}{\pi}\int_{\partial {\cal I}}\re^{-ik^{2}t}k{\cal 
F}(k)dk,\quad 0<t<T,
\label{invapp}
\ee
where $\partial {\cal I}$ denotes the boundary of the first quadrant 
${\cal I}$
of the complex $k$-plane, with counterclockwise orientation. 
It is interesting that small variations of 
elementary integrals such as the integral (\ref{dt1}), 
apparently have not  been investigated until now. Perhaps this is due 
to the fact that the analysis of the integral (\ref{dt1}), in contrast
to the analysis of the integral (\ref{ft1}), involves functions that are 
not analytic. Indeed, besides using the Fourier transform, there exists an 
alternative approoach for 
inverting (\ref{ft1}), which is based on the spectral analysis of the 
ODE
\be
\mu_{t}(t,k)+ik^{2}\mu(t,k)=kf(t),\quad 0<t<T, \;\;k\in\mathbb C.
\label{odeapp}
\ee
The crucial difference between this equation and the ODE 
(\ref{ode1}) (which is associated with the integral (\ref{dt1})) is 
that whereas there exists a solution $\mu(t,k)$ of equation 
(\ref{odeapp}) which is {\em sectionally analytic}, the solution of 
(\ref{ode1}) involves $\mu_{3}(t,k,\bar k)$ for which $\partial 
\mu_{3}/\partial \bar k\neq 0$.

The method used in this paper for inverting the integral (\ref{dt1}) 
has its origin in the paper \cite{fgel}, where it was 
emphasised that techniques developed for the solution of integrable 
nonlinear PDEs provide a new method for constructing integral 
transforms pairs. In particular, it was shown in \cite{fgel} that the 
spectral analysis of the ODE
$$
\mu_{x}(x,k)-ik\mu(x,k)=q(x), \quad x\in\mathbb R, \;\;k\in \mathbb C,
$$
yields the Fourier transform pair. The first nontrivial application 
of this method appeared in the work of R. Novikov \cite{nov}, who was able to 
invert the attenuated Radon transform by using a simple extension of 
a novel derivation of the Radon transform obtained in \cite{fnov} using 
the methodology of \cite{fgel}. 
It appears that our work, which yields the inversion of a large class 
fo integrals,  presents a second nontrivial application of 
the method of \cite{fgel}. These integrals are precisely the ones that characterise 
the Dirichlet to Neumann map for moving initial-boundary value problems. 
Given the simple form of these integrals, it is 
natural to expect that they may appear in other applications. 
The inversion of of the integral characterising the Dirichlet to 
Neumann map for the heat equation is presented in \cite{fokdel}.

The spectral analysis of equation (\ref{dt1}) was first carried out 
in \cite{prl, fpel} where the particular solutions $\mu_{1}$, 
$\mu_{2}$ and $\mu_{3}$ (see equations (\ref{mu1}), (\ref{mu2}), (\ref{mu3})) were 
introduced. 
However, in \cite{prl,fpel} the inversion formula for $f(t)$ was left 
in terms of a two-dimensional integral, and thus it did {\em not} 
provide an effective way of constructing $f(t)$. The crucial new development 
presented here is the understanding that the double integral can be 
expressed in terms of single integrals, which in turn yield a Volterra 
integral equation for $f(t)$.

\section*{Acknowledgments}
The authors gratefully acknowledge financial support for this research.
ASF was supported by a grant from EPSRC, while BP was supported by a Research 
Fellowship of the 
Leverhulme Foundation.

\end{document}